%% file: main.tex
\documentclass[11pt]{article}

\usepackage[margin=1in]{geometry}
\usepackage[utf8]{inputenc}
\usepackage{amsmath,amssymb,amsfonts}
\usepackage{graphicx}
\usepackage{textcomp}
\usepackage{booktabs}
\usepackage{flafter}
\usepackage{placeins}
\usepackage[font=small,labelfont=bf]{caption}
\usepackage[numbers,sort&compress]{natbib}
\usepackage{url}
\usepackage[hidelinks]{hyperref}

\graphicspath{{figures/}}
\providecommand{\tightlist}{%
  \setlength{\itemsep}{0pt}\setlength{\parskip}{0pt}}

\title{Circuit-free Cardiovascular Monitoring via Smartphone-readable\\
Skin-interfaced Nanophotonic Films}

\author{%
\parbox{0.96\textwidth}{\centering
Torjus Steffensen$^{1}$,
Vegar Stubberud$^{2}$,
Emilie S. Hennisen$^{3}$,
Eryk T. Winogrocki$^{3}$,\\
Julia L\"ovgren$^{3}$,
Arthur G. S. Torvund$^{3}$,
H\r{a}vard Vestad$^{2}$,
Nils K. Skj{\ae}rvold$^{1,4}$,\\
Martin Steinert$^{2}$, and
Angelos Xomalis$^{3,*}$\\[0.9em]
{\small
$^{1}$Department of Circulation and Medical Imaging, Norwegian University of Science and Technology, Trondheim, Norway\\
$^{2}$Department of Mechanical and Industrial Engineering, Norwegian University of Science and Technology, Trondheim, Norway\\
$^{3}$Department of Electronic Systems, Norwegian University of Science and Technology, Trondheim, Norway\\
$^{4}$St. Olav's University Hospital, Trondheim, Norway\\[0.7em]
$^*$Corresponding author: \href{mailto:angelos.xomalis@ntnu.no}{\texttt{angelos.xomalis@ntnu.no}}}
}%
}

\date{}

\begin{document}
\maketitle

\begin{abstract}
Nanoscale mechano-optical surfaces enable electronics-free strain sensing, attractive for skin-interfaced devices, yet reported implementations require laser/spectrometer interrogation, negating this advantage. Here, we report electrically passive mechanical transduction of arterial pulsation into diffractive colour shifts read by unmodified smartphone cameras, enabled by a dual-function monolithic poly-dimethylsiloxane (PDMS) film. Using large-area double-sided nanoimprinting, we achieve a strain-sensitive nanophotonic surface on one face of the film and a bio-inspired 3D structural adhesive on the other. We measure strain-dependent optical response and reproduce it in colour-mixing optical simulations. In uniaxial cyclic loading tests, 2\% strain produces a 9\% RGB-intensity modulation, stable over 1000 cycles. Further, 3D structuring improves adhesive shear strength by 65\% on skin over flat PDMS. Hand-held smartphone recordings in humans (n=13) resolve sub-beat hemodynamics in agreement with clinical reference (per-beat waveform \(\rho\)=0.94 \(\pm\) 0.03), exceeding established non-invasive techniques such as active reflectance photoplethysmography (PPG), imaging PPG, and piezoelectric pulse-force sensors in simultaneous recordings. Importantly, mechanical transduction at the elastomer--air interface presents an optical cardiovascular monitoring approach agnostic to dermal-melanin, a known PPG confounder. Together, these advances establish camera-readable mechanochromic elastomers as versatile materials platforms for wearable cardiovascular monitoring, point-of-care diagnostics, and electronics-free human-machine interfaces.
\end{abstract}


\input{body}

\input{references}
\end{document}

%% file: body.tex
\hypertarget{introduction}{%
\section{Introduction}\label{introduction}}

Continuous and non-invasive physiological monitoring increasingly enables timely diagnosis and intervention without compromising patient comfort or mobility. There is strong interest in wireless and minimal-contact solutions, yet wearable physiological sensing remains dominated by transduction architectures requiring on-body electronics: resistive, capacitive, piezoelectric, and electrochemical \citep{ref1,ref2,ref3}. This conversion can introduce artifacts inherent to the underlying physical process such as exhibited by piezoresistive devices \citep{ref1,ref4,ref5,ref6} and increase practical device complexity by requiring on-device electronics or power sources such as batteries, external wires, or energy harvesting. A direct, electrical transduction-free measurement approach avoiding these restrictions will have a great advantage.

Emerging remote sensing methods attempt to address these constraints \citep{ref2,ref7}. Fully non-contact monitoring using radiative elements-based technologies can be effective for vital sign and respiratory tracking but typically lack the spatial resolution necessary for precise characterization of subtle cardiovascular movements unless placed close to the body \citep{ref8}.

Metamaterials and photonic crystals, engineered nanostructures with tailored optical properties \citep{ref9}, have recently shown great promise in sensing applications due to their exceptional detectivity of chemical and physical changes in their proximity. Briefly, plasmonic sensing platforms leveraging surface-enhanced Raman scattering (SERS)\citep{ref10} and fluorescence have been used to detect chemical markers in biofluids for applications like metabolic monitoring and diagnostics with high chemical specificity and low detection limits \citep{ref11,ref12}. Topological metamaterial-based wearable sensors capable of detecting vital signs using radiofrequency interrogators have been demonstrated \citep{ref13}.

Photonic wearable sensors specifically translate physiological signals into optical changes that can be measured remotely, offering an alternative to the electrochemical, resistive, and transmittive transduction architectures that dominate the wearable healthcare landscape \citep{ref11,ref14}. The most established category of wearable photonic sensors is based on light-tissue-interactions, such as photoplethysmography (PPG), by far the dominant optical cardiovascular modality in clinical and research use. PPG techniques may be skin-interfaced, or non-contact such as in remote or imaging PPG, relying on reflected light. Imaging PPG methods are desirable due to the singularly simple instrumentation involved but are less robust in uncontrolled environments \citep{ref15}. Because the optical path passes through the dermis, PPG inherits a documented sensitivity to dermal melanin variability across pigmentation \citep{ref16,ref17}, an issue receiving increasing regulatory scrutiny \citep{ref18}, and typically recovers lower-frequency components of the pulse wave that survive tissue scattering \citep{ref7,ref15}.

Beyond PPG, non-tissue-interacting photonic methods based on structural-photonic platforms have been demonstrated. Crucially, such applications can be probed remotely without on-device electrical conversion \citep{ref19}. Demonstrated approaches include fiber optic strain sensors \citep{ref13,ref20,ref21}, pressure sensitive 3D printed woodpile lattices \citep{ref19}, plasmonic metasurfaces \citep{ref10,ref12,ref22} and colloidal dispersion in polymer matrices by measuring strain-induced spectral shifts \citep{ref22,ref23,ref24}. Common to all these systems is the need for complex optics and external interrogators, i.e.~spectrometers, lasers, or infrared optics. The benefits of the ``wireless'' readout are thus often eliminated in practice. The potential for truly ``wireless'' and circuit-free acquisition of physiological mechanical signals, such as those produced by arterial pulsation, respiration, or tissue deformation parameters crucial to non-invasive cardiovascular diagnostics, remains greatly underexplored.

Intriguingly, nanostructured poly-dimethylsiloxane (PDMS) films and RGB camera systems have been demonstrated for magnetic field detection pointing to applications in circuit-less monitoring in medical imaging and drug transport \citep{ref25,ref26}. More recently, Arief et al used visible spectrum diffractive PDMS optics to characterize strain behavior in an electrically transducing device\citep{ref27}. Together, these advances suggest a wearable architecture in which a commodity camera, such as a smartphone, interrogates a passive elastomeric skin-interfaced transducer. A simple, disposable, and potentially biodegradable film remains on the body, while power and electronic complexity are moved to a device the user already owns. The on-skin component is battery-free and contains no critical raw materials, avoids delamination of rigid electronics from skin, and due to the ubiquity of smartphones may reduce barriers to cardiovascular screening.

In this study, we demonstrate the application of mechanochromic sensing using a flexible dielectric nanopatterned surface to continuously record cardiovascular phenomena in humans using unmodified smartphone cameras. We utilize large-area double-sided nanoimprinting techniques to provide optically active and structural adhesion functionality simultaneously. Next, we evaluate the optical and mechanical response of nanophotonic meta-grating sensors, showing strong linearity (R\textsuperscript{2} = 0.93) over more than 1000 load cycles. We test the biomimetic adhesion surface in peel- and pull-off experiments on skin and glass substrates, showing a significant improvement in substrate adhesion. In 13 healthy volunteers, we recover pulsatile waveforms with mean per-beat Pearson correlation of 0.94 (SD 0.043) and identify benign arrhythmic events. For clarity, we perform a detailed comparison against established electro-optical and piezoelectric pulse sensors, validating our technique's fidelity for cardiovascular monitoring. Capturing such feature-rich cardiovascular recordings gives access to subtle hemodynamic information, an evaluator of e.g.~arterial stiffness \citep{ref28,ref29,ref30}, an indicator of cardiovascular health \citep{ref31,ref32} and may improve performance of machine learning and deep learning-based estimators on the basis of the richer feature content \citep{ref4}.

\begin{figure}[tbp]
\hypertarget{fig:1}{%
\centering
\includegraphics[width=\linewidth,height=0.68\textheight,keepaspectratio]{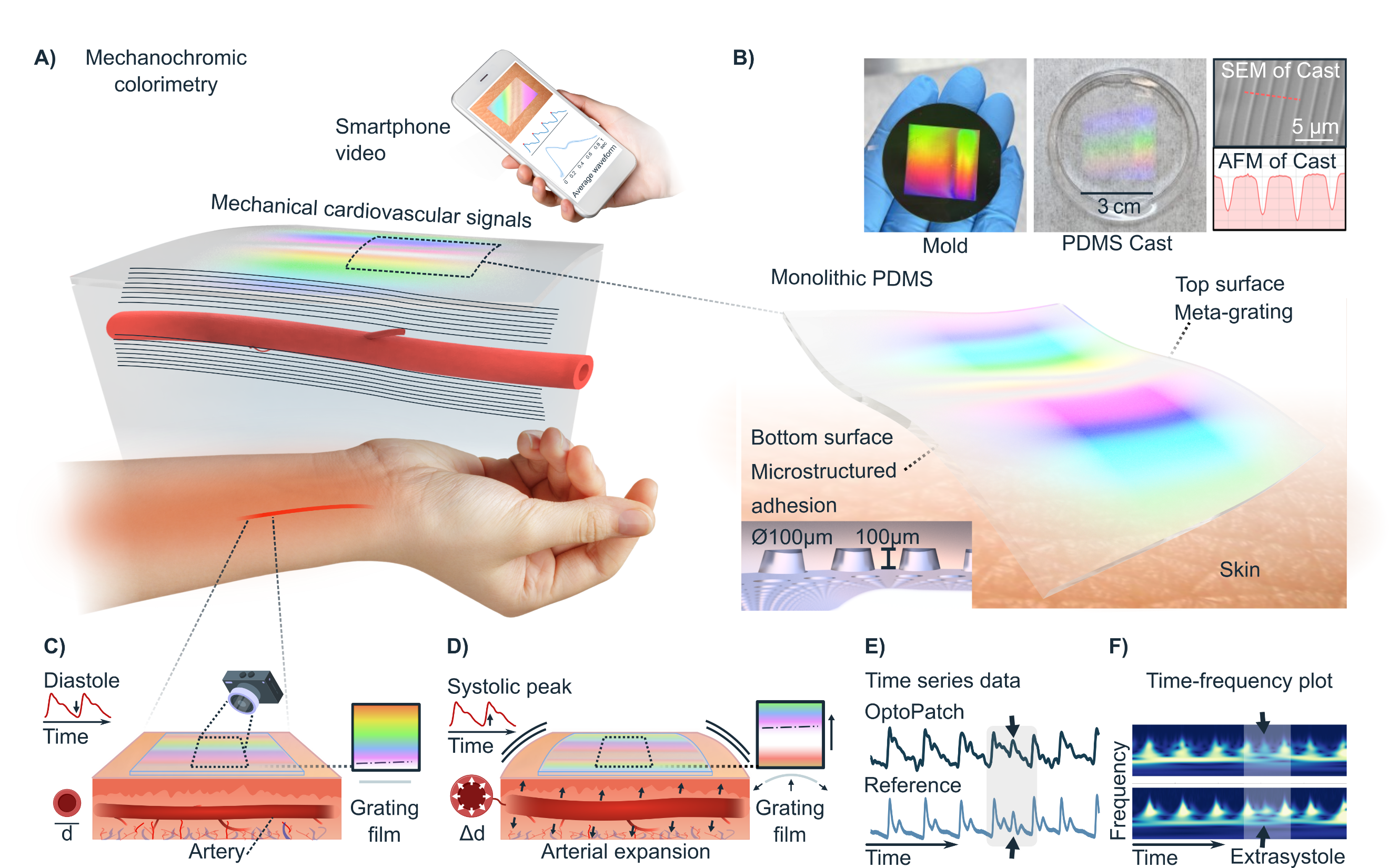}
\caption{\textbf{Concept of circuit-free cardiovascular monitoring via mechanochromic colorimetry. A)} OptoPatch operating principle: a flexible nanophotonic dielectric meta-grating film adhered to the skin above the radial artery translates arterial expansion into diffraction colour shifts recorded by a smartphone camera. \textbf{B)} Device structure. Top: photographs of the nanofabricated mold (left) and resulting poly-dimethylsiloxane (PDMS) cast (right) with scanning electron microscopy (SEM) and atomic force microscopy (AFM) insets confirming structure quality. Bottom: schematic of the monolithic PDMS film showing the metasurface on the top surface and the 3D microstructured suction-cup adhesive on the bottom surface (cavity diameter \O{}100 \textmu{}m, depth 100 \textmu{}m), with the device conforming to skin directly. \textbf{C--D)} Working principle. During diastole (C), the artery is at baseline diameter, and the metasurface patch is relaxed; during systole (D), arterial expansion deforms the overlying skin-interfaced film, shifting diffraction angles and the observed colour pattern relative to the camera pixels. \textbf{E)} Representative time-series data from OptoPatch (top) and a continuous non-invasive blood pressure (cNIBP, Finapres) reference (bottom) during an arrhythmic event (arrows). \textbf{F)} Normalized continuous wavelet transform scalograms of the segment in (E) from OptoPatch (top) and NIBP (bottom), with arrows indicating the arrhythmic event.}\label{fig:1}
}
\end{figure}

\hypertarget{results-and-discussion}{%
\section{Results and discussion}\label{results-and-discussion}}

\hypertarget{monolithic-patch-fabrication}{%
\subsection{Monolithic patch fabrication}\label{monolithic-patch-fabrication}}

To obtain the desired mechanochromic behaviour 1-D gratings were designed with a range of periodicities (600 and 2100 nm). We fabricated nanostructured negative molds via two complementary methods. Maskless lithography produced large-area periodic structures (3\(\times\)3 cm\textsuperscript{2}) with controlled periodicity 2100 nm, highlighting scalability through a mature fabrication process (Figure \ref{fig:1}B, top right). In parallel, we used Two-Photon Polymerization (2PP) (NanoOne, UpNano) to create smaller, highly customizable structures (1\(\times\)0.5 cm\textsuperscript{2}) at a periodicity of 600 nm. This approach demonstrated potential for rapid prototyping of more complex 3D geometries. PDMS was spin-coated over the mold surface, creating a reverse (single-step cast) of the mold grating resulting in a flexible polymeric film (Figure S1). Visible diffraction patterns under ambient light confirmed successful transfer of the grating structures. Additionally, scanning electron (SEM) and atomic force microscopy (AFM) of the cast elastomer confirmed good surface quality (Figure \ref{fig:1}B). Individual photonic and adhesive master molds were successfully re-used multiple times (6-10) outside the cleanroom environment without noticeable degradation (Figure S10), consistent with previous reports demonstrating the reusability of the master after repeated PDMS casting.

\begin{figure}[tbp]
\hypertarget{fig:2}{%
\centering
\includegraphics[width=\linewidth,height=0.68\textheight,keepaspectratio]{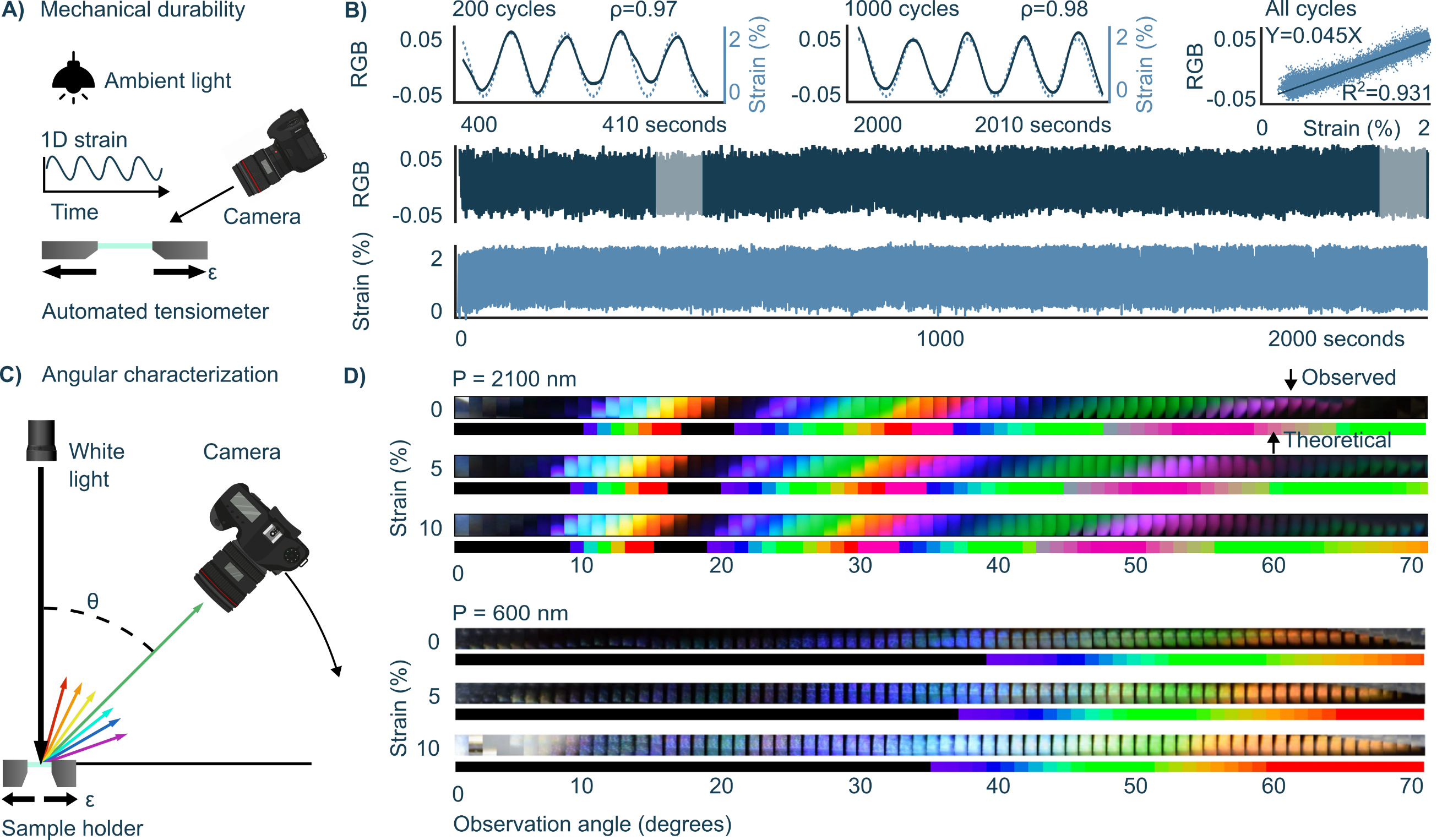}
\caption{\textbf{Optics and mechanics of photonic sensor. A)} Mechanical durability test configuration: a uniaxial tensiometer applies cyclic strain to the OptoPatch sample under ambient illumination while a camera records the surface. \textbf{B)} Durability results. Top left and top center: relative scaled RGB pixel intensity (range 0-1) signal (solid line) excerpts after 200 and 1000 strain cycles (dashed line), showing little appreciable drift or amplitude variation. Top right: RGB pixel intensity versus strain cycles for all data, showing good linearity across the range. Middle: RGB pixel intensity over the recorded video. A scalar baseline has been subtracted to give the recording zero mean. Bottom: applied strain profile over the full 2000 sec test. \textbf{C)} Goniometric characterization setup: an uncollimated white-light source illuminates the sample at approximately normal incidence and a digital RGB camera records reflected colours across observation angles \(\theta\) under applied uniaxial strain \(\varepsilon\). \textbf{D)} Angular colour response at 0, 5, and 10\% strain for gratings with periodicities of 2100 nm (top) and 600 nm (bottom). For each periodicity, observed and computationally simulated colour strips are shown, demonstrating correspondence between experiment and theory (CIE 1931 colour model; see Supplementary Material).}\label{fig:2}
}
\end{figure}

The capability of the structures to diffract light into multiple orders was further studied by measuring the laser beam power in the reflected diffraction orders (Figure S4). Importantly for upscaling and rapid prototyping, once the molds were created all sensor devices were fabricated outside cleanroom conditions.

\hypertarget{optical-benchtop-characterization}{%
\subsection{Optical benchtop characterization}\label{optical-benchtop-characterization}}

Figure \ref{fig:2} illustrates behavior of the OptoPatch in benchtop conditions. To establish mechanical durability, an automated tensiometer strained a 3.5\(\times\)0.98 cm (3.0 cm free length) sample device of thickness 340 \textmu{}m to a physiologically realistic range of uniaxial strain at a 0.47mm peak-to-peak axial displacement (\(\approx\)2\% nominal peak-to-peak strain). A smartphone camera (iPhone 14, Apple, USA) on a tripod recorded video of the surface of the sample.

We included \(\sim\)2000 seconds of continuous cycling at 0.5Hz (n=1009 complete cycles) for analysis (Figure \ref{fig:2}B). There was no notable drift or variation in signal amplitude indicating good stability over time. Using the same data, we estimated transduction sensitivity with a smartphone camera in a stable and well-lit benchtop environment. Per-cycle peak-to-peak amplitude ratio gave a sensitivity of 0.134 (\(\pm\) 0.018) per mm displacement (units between 0-1, scaled RGB intensity) or equivalently 0.046 (\(\pm\) 0.006) per \% strain. Signal-to-noise ratio was estimated at 10.88 dB (Figure S7). The observed peak-to-peak modulation (covering \(\sim\)9.2\% intensity range from 0 to 2\% strain) sits well above smartphone-camera noise floors. This range is consistent with observed \emph{in vivo} fidelity. We also conducted extended-cycle durability testing to assess the long-term durability of the sensors, observing no degradation in signal quality at \textgreater43,000 cycles nor loss of profile geometry at 18,000 cycles (Figure S8--S9). There was no meaningful difference between different casts under similar loading (Figure S11), demonstrating replicability. Further readout improvements are likely to come from imaging-side optimisation and the spatial processing pipeline, particularly efforts to track high-contrast sample regions across the recording.

To quantify the sensor's optical strain response and practical viewing range, we conducted a goniometric study (Figure \ref{fig:2}C--D). In a uniaxial tensiometer horizontally mounted samples were gently pre-tensioned to ensure flatness. An uncollimated white-light source illuminated the samples at approximately normal incidence, and a digital RGB camera recorded reflected colors across viewing angles from 5 to 70\(^\circ\) (Figure \ref{fig:2}D). We applied controlled tensile strains of 0, 5, and 10\% relative to baseline.

Figure \ref{fig:2}D shows clear, progressive shifts in diffraction order onset and color response correlated with increasing strain and viewing angle. Further, we show results from computational simulations of the reflected diffraction orders with corresponding periodicity and observation angle. As several diffraction orders are expected to overlap, colors were mixed according to the CIE 1931 color model (Figure S3). We note correspondence between theoretical predictions and empirical observation, indicating the possibility to use simulations to tailor devices with high sensitivity to both flexion (angle change) and strain (periodicity change) in a desired regime.

\hypertarget{microadhesion-characterization}{%
\subsection{Microadhesion characterization}\label{microadhesion-characterization}}

\begin{figure}[tbp]
\hypertarget{fig:3}{%
\centering
\includegraphics[width=\linewidth,height=0.68\textheight,keepaspectratio]{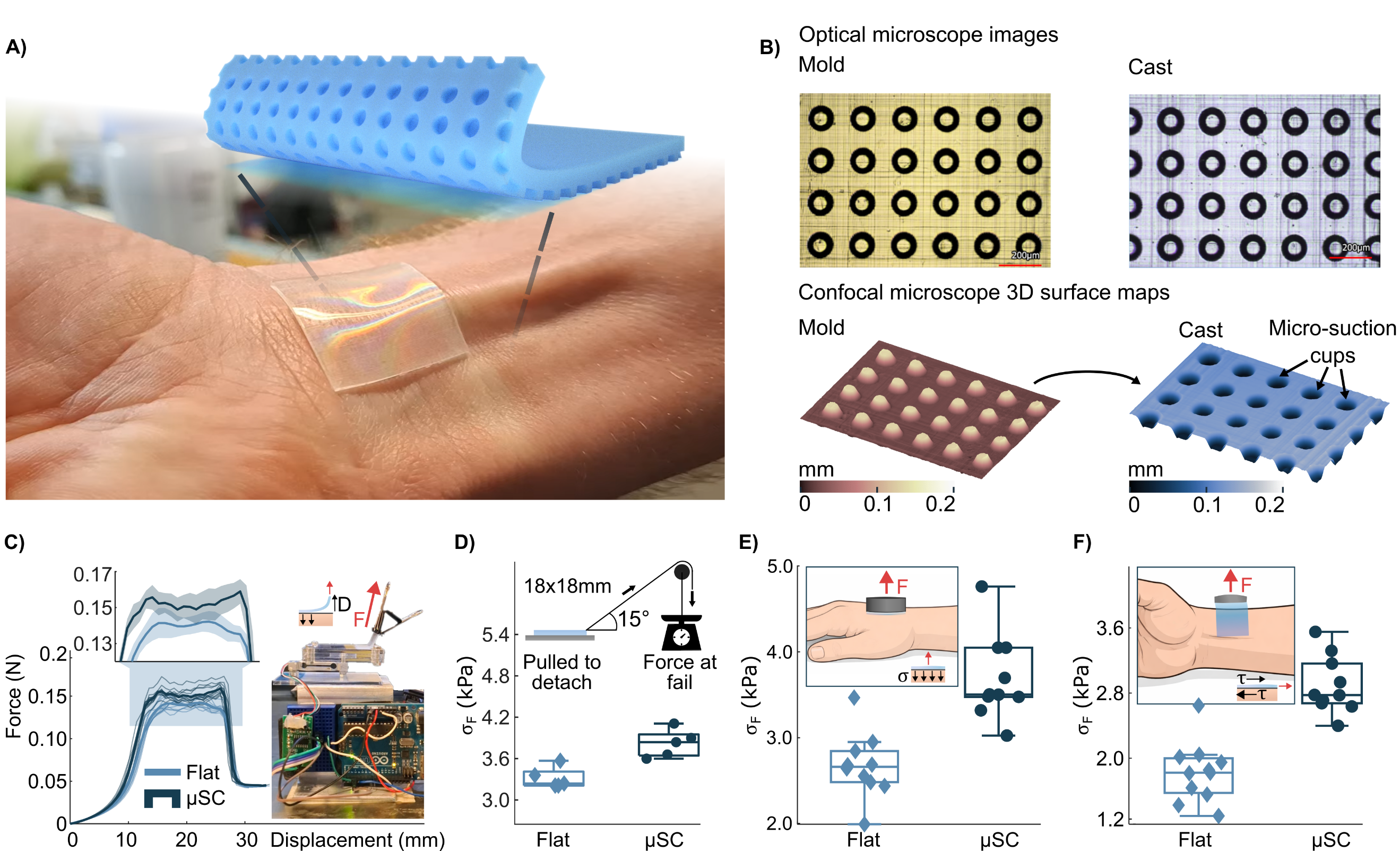}
\caption{\textbf{Functional adhesion design of photonic sensor. A)} Photograph of the OptoPatch adhered to human skin, showing the monolithic device with the micro-suction cup (\textmu{}SC) backing layer (bottom surface) conformally contacting the skin and the iridescent metasurface (top). \textbf{B)} Fabrication fidelity. Top row: optical micrographs of the Two-Photon-Polymerization (2PP)-printed mold (left) and replicated PDMS cast (right); scale bars 200 \textmu{}m. Bottom row: confocal microscopy height maps of the mold (left, negative pattern) and PDMS cast (right, positive replica). \textbf{C)} Peel test on a flat and rigid substrate. Force--displacement curves for flat PDMS and \textmu{}SC surfaces under vertical edge-initiated loading. Inset highlights the steady-state force regime; photograph and schematic show the experimental setup and loading geometry. \textbf{D)} Angled shear pull test on glass substrates (18\(\times\)18 mm samples, 15\(^\circ\) loading angle), reporting force at failure (and converted to stress \(\sigma_F\)) for flat and \textmu{}SC surfaces. \textbf{E)} Normal pull-off adhesion on human skin, quantified by \(\sigma_F\). Inset illustrates the applied loading geometry. \textbf{F)} Shear adhesion on skin. Inset illustrates the applied loading geometry. In all panels, individual data points are shown alongside boxplots summarising the distributions.}\label{fig:3}
}
\end{figure}

Figure \ref{fig:3} shows the OptoPatch \emph{in situ} above the wrist adhered with the micro-suction cup (\textmu{}SC) surface (Figure \ref{fig:3}A) alongside an illustration of the 3D adhesion geometry (not to scale). The design of the skin adhesion consists of inverted pillars modelled after a demonstrated \textmu{}SC geometry\citep{ref33}, fabricated using a positive mold (Figure \ref{fig:3}B, top left) printed using 2PP (top right). Successful transfer of the inverted pillar pattern is confirmed by comparing the mold and produced PDMS cast geometry in confocal microscopy (bottom). To assess adhesion of the \textmu{}SC surface, we conducted peel- and pull-off tests (Figure \ref{fig:3}C--F)\citep{ref33,ref34}. Figure \ref{fig:3}C shows edge-peel pull-off failure versus a flat PDMS surface alongside a photograph insert of the measurement setup. To perform a peel, a clamp is attached to the device edge connected to an Arduino-driven automated load cell. A computer-controlled vertical stage provides a constant vertical velocity, yielding a measurement of the force versus displacement from onset of peel. Similar setups were used for normal pull and shear measurements. In flat, dry adhesion on a smooth acrylic plane the \textmu{}SC surface increased peeling force by about 15\%, demonstrating an active suction effect. The \textmu{}SC structure improved adhesion compared to flat PDMS across all load scenarios, ranging from 20\% in flat surface shear (Figure \ref{fig:3}D) to a 65\% improvement in shear adhesion on skin (Figure \ref{fig:3}F), confirming skin adhesion effect and feasibility of the PDMS monolithic device concept. Similar \textmu{}SC devices have demonstrated long-term adhesion under exposure to sweat and moisture, as well as during robust arm motion \citep{ref33,ref35}. Prior to in vivo test we perform device benchmarking utilizing a wrist phantom module (Figure S5).

\hypertarget{cardiovascular-monitoring-in-healthy-volunteers-using-nanophotonic-patches}{%
\subsection{Cardiovascular monitoring in healthy volunteers using nanophotonic patches}\label{cardiovascular-monitoring-in-healthy-volunteers-using-nanophotonic-patches}}

\begin{figure}[tbp]
\hypertarget{fig:4}{%
\centering
\includegraphics[width=\linewidth,height=0.68\textheight,keepaspectratio]{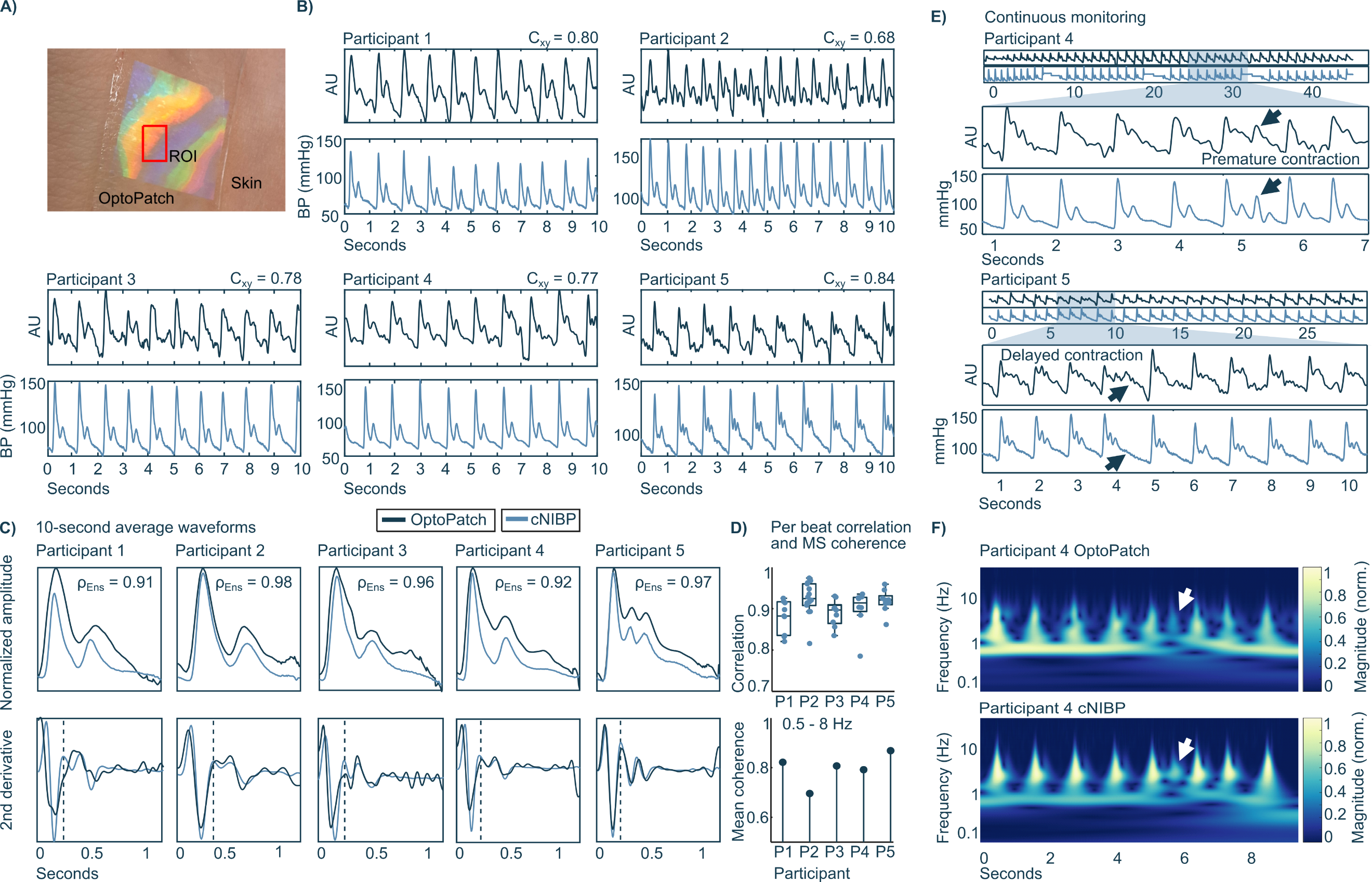}
\caption{\textbf{\emph{In vivo} cardiovascular recordings via wearable photonic sensor. A)} Video frame of the OptoPatch on human skin, with the region of interest (ROI) used for signal extraction indicated. \textbf{B)} Synchronized 10-second recordings from five participants showing extracted OptoPatch signals (dark blue traces, arbitrary units) of varying quality and simultaneously recorded cNIBP (light blue, mmHg). Mean magnitude-squared coherence (\(C_{xy}\), 0.5--8 Hz) is reported for each participant. \textbf{C)} Ensemble-averaged waveforms. Top: normalized OptoPatch (dark blue) and cNIBP (light blue) pulse waveforms averaged over the 10-second recording, with ensemble Pearson correlation (\(\rho_{\mathrm{Ens}}\)). Bottom: second derivatives of the averaged waveforms; dashed lines indicate the onset of the reflected systolic wave, identifiable in participants 2--5 for OptoPatch. \textbf{D)} Quantitative agreement metrics. Top: per-beat Pearson correlation between OptoPatch and cNIBP for each participant (individual beats shown with boxplot summary). Bottom: mean magnitude-squared coherence (0.5--8 Hz band) for each participant. \textbf{E)} Detection of benign arrhythmic events during continuous monitoring. Top: participant 4 recording (\(\sim\)45 s) with zoomed segment showing a premature contraction (arrows) visible in both OptoPatch and cNIBP traces. Bottom: participant 5 recording showing a delayed contraction (arrows) identifiable in both channels. \textbf{F)} Normalized wavelet scalograms for participant 4 from OptoPatch (top) and cNIBP (bottom). White arrows indicate the extrasystole event, clearly resolved in both channels up to \(\sim\)10 Hz.}\label{fig:4}
}
\end{figure}

We performed a study in healthy volunteers (n=13) at rest as a proof-of-concept demonstration. Results are shown in Figure \ref{fig:4}. During operation, OptoPatch sat on the skin above the radial artery of the non-dominant hand and was illuminated by a ceiling panel light (Figure \ref{fig:4}A). Video was recorded from a hand-held smartphone camera, using two different mid-range smartphone models (Galaxy A55 5G, Samsung Electronics; iPhone 14, Apple) using the default camera app with automatic focus, exposure, and white balance. No attempt was made to ensure controlled lighting conditions in either of two rooms used for recording (Figure S12). To provide a baseline comparison, we recorded continuous non-invasive blood pressure (cNIBP) in the finger artery of the same limb simultaneously with a volume clamp device (NIBP-Nano, ADInstruments). Finger volume-clamp waveforms show well-established agreement with invasive arterial pressure and close amplitude- and frequency-domain similarity to radial tonometry, supporting their use as an appropriate comparator for radial artery waveform morphology \citep{ref36,ref37,ref38}. Because the two modalities use fundamentally different transduction pathways absolute signal magnitudes are not comparable by construction directly. The physiologically informative correspondences are instead the morphology of the waveform (Figure \ref{fig:4}C) and the relative timing of its fiducial feature. Figure \ref{fig:4}B presents synchronized pulse trains from five participants extracted from the OptoPatch (dark blue line) and NIBP (light blue). Waveform quality varied across participants, likely due to uneven adhesion, but in all cases we recovered pulsatile signals from the video recordings with good accuracy. Figure \ref{fig:4}C shows ensemble averages created by segmenting the pulse train into individual cycles and averaging sequential beats. These waveforms show clear similarity between OptoPatch and NIBP.

In participants 2-5 the reflected pressure wave is readily identified by the distinctive deflection of the second derivative (dark blue line, Figure \ref{fig:4}C bottom) \citep{ref28,ref39}. The ability to identify subtle features such as the inflection point (dashed line) associated with the reflected pulse wave enables pulse contour analysis and assessment of relative pulse amplification in the peripheral arteries. Importantly, we achieve this fidelity of the recovered waveforms under unstandardized lighting and handheld imaging, underscoring the robustness of our optical transduction mechanism. Further, the measurements highlight tentatively identified benign arrhythmic events for participant 4 and 5, reflected in both cNIBP and OptoPatch (Figure \ref{fig:4}E black arrows). Participant 4 exhibits a sharply oscillating complex in late diastole, identified as a noncompensatory premature contraction. Participant 5 shows an extended diastolic period between beats, identified as a delayed contraction. To extract more information, we plot the time-frequency representation with a continuous wavelet transform (CWT) of the segment from participant 4 (Figure \ref{fig:4}E). While the main heartbeat activity is around 1 Hz, there is rich information in the range up to 10 Hz. Here, OptoPatch (top, Figure \ref{fig:4}F) clearly resolves an extrasystole event (white arrow) in great agreement with NIBP (bottom).

\begin{figure}[tbp]
\hypertarget{fig:5}{%
\centering
\includegraphics[width=\linewidth,height=0.68\textheight,keepaspectratio]{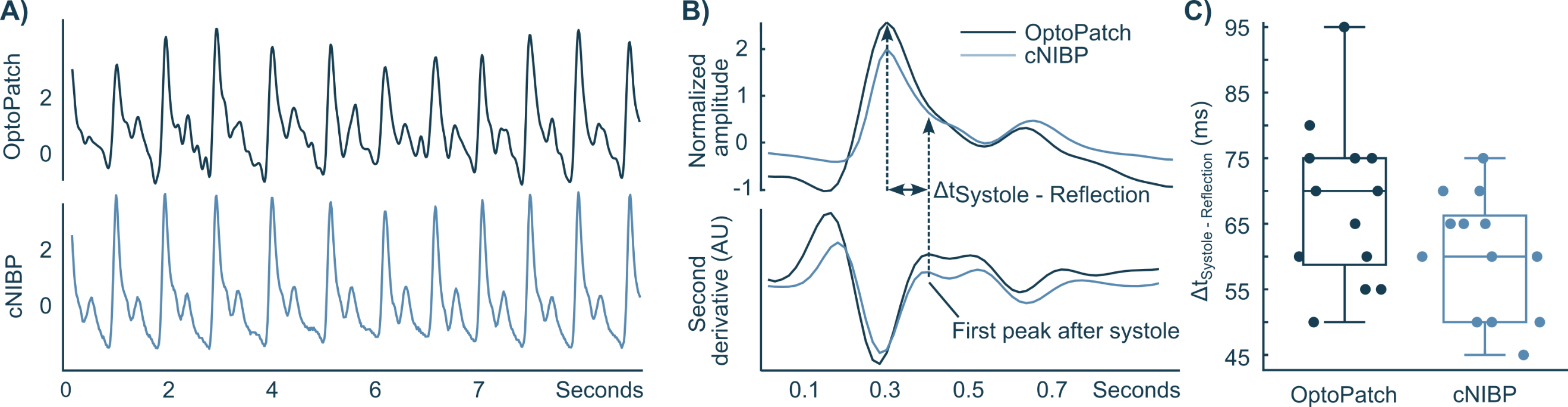}
\caption{\textbf{Sub-beat hemodynamic timing.} \textbf{A)} Synchronized time-series from: OptoPatch colorimetric signal (top) and continuous non-invasive blood pressure reference (cNIBP, bottom), recorded via handheld smartphone at 60 FPS under uncontrolled ceiling-panel lighting. \textbf{B)} Ensemble-averaged waveform (top) and second derivative (bottom) for OptoPatch (dark) and cNIBP (light), with the systolic-to-reflected-wave interval (\(\Delta t_{\mathrm{Systole--Reflection}}\)) indicated. The reflected wave onset is identified as the first peak of the second derivative following systole. \textbf{C)} Systolic-to-reflected-wave timing interval (\(\Delta t_{\mathrm{Systole--Reflection}}\)) derived independently from OptoPatch and cNIBP derivative analysis across the cohort (n = 13). The camera sample rate is \(\sim\)16.7 ms per frame (upsampled to 5ms per sample via cubic interpolation), limiting resolution. Distributional overlap nevertheless indicates that sub-beat hemodynamic timing is accessible from mechanochromic transduction. Box: median and interquartile range; whiskers: full data range; dots: individual participants.}\label{fig:5}
}
\end{figure}

To probe whether OptoPatch resolves physiologically meaningful inter-beat timing information, we further quantified the interval between the systolic peak and the onset of the reflected wave (or the inflection point), as \(\Delta T_{\mathrm{Systole--Reflection}}\), independently in the OptoPatch and NIBP signals across the cohort (n = 13) (Figure \ref{fig:5}). The reflected-wave onset was defined as the first peak of the second derivative following systolic peak \citep{ref28,ref39}. Figure \ref{fig:5}A shows representative synchronized time-series from the two modalities; Figure \ref{fig:5}B shows the ensemble-averaged waveforms and their second derivatives with the \(\Delta T\) interval annotated. The reflected-wave feature was identifiable in both signals in 12 of 13 participants, ranging from clearly present to identifiable only via derivative analysis. In 1 participant, the reflected wave was identifiable in the NIBP signal but not the OptoPatch. In 12 participants, the per-participant \(\Delta T\) distributions derived from OptoPatch and from NIBP overlap substantially (Figure \ref{fig:5}C). The camera sample rate (\(\sim\)16.7 ms per frame, upsampled to 5 ms via cubic interpolation) places a lower bound on timing resolution that must be considered when interpreting these distributions. The distribution overlap establishes that sub-beat hemodynamic timing is recoverable from a passive, circuit-free mechanochromic film read by a consumer smartphone across participants. We evaluated the devices at a framerate of 60 FPS as it is a widely supported capability in commodity smartphones and therefore readily deployable. We note that as higher video frame rates become available (e.g.~120 FPS) these may better capture rapid transients, intra-beat dynamics, and subtle arrhythmic features.

\hypertarget{comparison-to-established-sensor-technologies}{%
\subsection{Comparison to established sensor technologies}\label{comparison-to-established-sensor-technologies}}

\begin{figure}[tbp]
\hypertarget{fig:6}{%
\centering
\includegraphics[width=\linewidth,height=0.68\textheight,keepaspectratio]{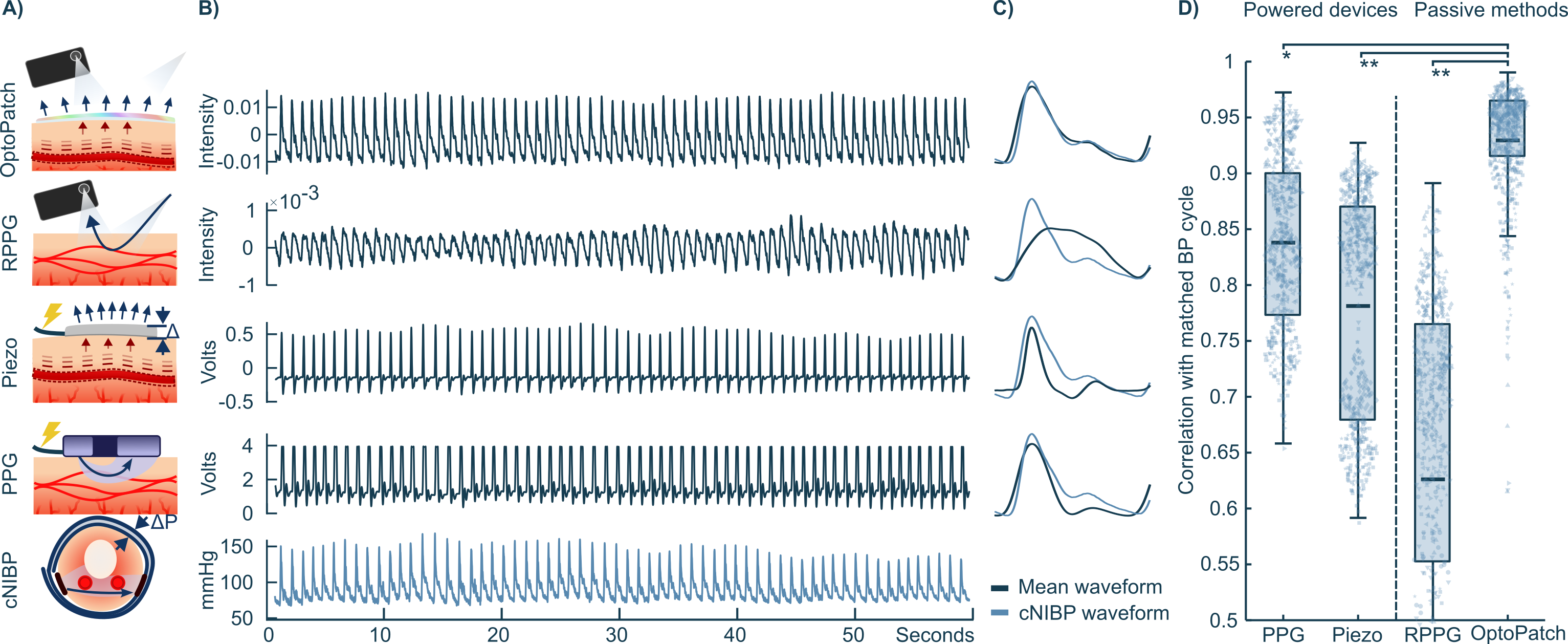}
\caption{\textbf{Comparison with other non-invasive sensing methods.} \textbf{A)} Schematic illustration of the operation of OptoPatch and three relevant comparable cardiovascular sensors. From top: OptoPatch performs passive optical sensing of mechanical deformation; remote-PPG (RPPG) performs passive optical sensing of light-tissue interaction; piezoelectric force sensors generate electrical potential in response to mechanical deformation; reflectance PPG actively senses light-tissue interaction. cNIBP is recorded via finger cuff volume clamp. \textbf{B)} Example concurrent recording showing, from top to bottom, OptoPatch, RPPG, piezoelectric sensor, reflectance PPG, and comparator cNIBP. All signals bandpass-filtered and baseline subtracted. RPPG, PPG have been inverted. \textbf{C)} Averaged waveforms from B) showcasing morphological similarity with the BP waveform (light blue). \textbf{D)} Correlation of normalized individual beats across a dataset of 8 participants for whom comparator data was available. Boxes indicate inter-quartile range, whiskers denote 1.5 times IQR, individual points are beats. Solid horizontal bar is the mean of participant means. Asterisks indicate significant differences, *: p \textless{} 0.05, **: p \textless{} 0.01.}\label{fig:6}
}
\end{figure}

Further to prove the validity of our methodology, we compare the OptoPatch performance against established noninvasive sensors. We assess time-domain correlations with cNIBP (Figure \ref{fig:6}B, bottom) for all cases. Example waveforms constructed from assembling consecutive heartbeats are shown superimposed on the reference blood pressure (BP) waveform (Figure \ref{fig:6}C), indicating typical differences and similarities. We note that not every sensor method in the comparison directly measures a signal that is strictly proportional to the BP contour. For instance piezoelectric type sensors exhibit typical artifacts consequent of the time-varying excitation of the material along its thickness \citep{ref5}. The PPG signal, likewise, is a measurement of the blood volume pulse rather than pressure, and varies morphologically with sensor geometry (transmission versus reflectance) and measurement site (e.g.~fingertip versus wrist)\citep{ref40}, achieving higher similarity to the pressure pulse waveform in transmission mode at the fingertip\citep{ref41,ref42}. Here, a consumer-grade fingertip reflectance-type sensor was used for ease of synchronization with the other comparators due to its analog output. A transmission-type fingertip device would likely have yielded a somewhat higher mean correlation, though we note recordings clustered above 0.9 as well. In any case, as OptoPatch transduces mechanical deformation on the skin surface modulated by lumen distension induced by the pressure pulse, comparison to the BP contour gives some indication of device performance in context.

OptoPatch r was higher than RPPG (n = 8 participants; \(\Delta r\) = 0.304, 95\% CI {[}0.235, 0.398{]}; p \textless{} 0.001), Piezo (n = 8; \(\Delta r\) = 0.130, 95\% CI {[}0.085, 0.196{]}; p \textless{} 0.001), and PPG (n = 6; \(\Delta r\) = 0.061, 95\% CI {[}0.018, 0.126{]}; p = 0.031).

Notably, by this metric the material enables passive sensing on par or in some cases superior to active skin-contacting modalities (Figure \ref{fig:6}D), outperforming imaging RPPG.

Notably, because OptoPatch signals arise from diffraction at the free upper surface of the meta-grating film the transduction pathway is insensitive to light-skin interactions, including dermal melanin absorption. This is a structural distinction from PPG, whose reliance on tissue-transmitted light underlies its documented pigmentation-related performance variation\citep{ref16,ref17,ref18}. In our cohort, OptoPatch produced canonical arterial pulse waveforms in participants with varying skin pigmentation, including clearly resolved reflected waves and dicrotic notches (Figure S6), consistent with a surface-diffraction readout that does not optically interrogate the skin. This performance may support mechanochromic transduction as a path to inexpensive and equitable optical cardiovascular monitoring and this may concern future investigations. We note that variation in the thickness and mechanical properties of intervening soft tissue may affect the transmission of arterial motion to the skin surface, damping the pulse. Future work will characterize tissue--sensor coupling to investigate performance across broader populations.

\FloatBarrier

\hypertarget{conclusion}{%
\section{Conclusion}\label{conclusion}}

In summary, we show a wearable circuit- and fiber-optic free photonic sensor recovering arterial pulse-wave morphology from an electrically passive elastomer film interrogated by an unmodified smartphone camera. The device is a single PDMS film carrying two functionally distinct nanostructured surfaces produced in a single casting step, combining a mechano-optically transducing diffractive surface and a biomimetic structural adhesive for skin attachment in one contiguous material layer.

We performed extended mechanical and optical characterization, demonstrating stable linear mechanochromic response to physiologically relevant strains using commodity smartphone cameras. Goniometric mechano-optical characterization established strain-dependent optical response in agreement with computational simulation, and a wrist-mimicking phantom validated transduction prior to \emph{in vivo} testing. In thirteen healthy volunteers, OptoPatch yielded arterial pulse traces with high morphological agreement against cNIBP reference. Intra-beat hemodynamic events like wave reflection were identifiable in twelve of thirteen participants.

Unlike existing wearable photonic sensors, divided between RF-interrogated structural-materials platforms and laser- or spectrometer-interrogated photonic-materials platforms, OptoPatch combines passive optical transduction with an ambient consumer-grade readout instrument. Notably, in head-to-head comparison against active skin contact PPG (RPPG), imaging PPG, and a piezoelectric pulse force transducer, OptoPatch achieved superior per-beat morphology agreement with the cNIBP reference. Because transduction occurs at a free elastomer--air interface, the signal pathway is decoupled from light-tissue absorption and from the dermal-melanin variability that otherwise constrain all PPG-based sensors. We anticipate expansion to metasurface nano-optics together with developments in imaging-side processing will extend the system's tolerance to motion and ambient-light variability.

By coupling skin-interface nanophotonic elastomers with commodity-camera readout, this work provides a simple route to waveform-level cardiovascular sensing without embedded electronics or specialized interrogation hardware. This work positions camera-readable mechano-optical materials as broadly accessible measurement interfaces across photonics, soft matter, human-machine interaction, and biomedical instrumentation. Importantly, delivering high fidelity cardiovascular data will impact future investigations involving machine- and deep-learning methodologies for ubiquitous health.

\hypertarget{methods}{%
\section{Methods}\label{methods}}

\hypertarget{photolithographic-grating-molds}{%
\subsection{Photolithographic grating molds}\label{photolithographic-grating-molds}}

Silicon (Si) wafers were solvent-cleaned (acetone, isopropanol) and UV/ozone treated (3 min, 30 \(^\circ\)C), dehydrated (120 \(^\circ\)C, 6 min), HMDS-primed (20 min), then briefly dehydrated (110 \(^\circ\)C, 60 s). SPR700-1.0 was spin-coated (4000 rpm, 34 s), soft-baked (95 \(^\circ\)C, 60 s), exposed (115 mJ cm\(^{-2}\)), post-exposure baked (115 \(^\circ\)C, 60 s), and developed (ma-D 332-S, 60 s). A hard bake (130 \(^\circ\)C, 3 min) completed the patterning. PFOCTS was applied as an anti-stick monolayer. Gratings with periodicities of 1700 and 2100 nm were fabricated.

\hypertarget{two-photon-fabricated-gratings}{%
\subsection{Two-photon fabricated gratings}\label{two-photon-fabricated-gratings}}

Gratings with a 600 nm periodicity were fabricated by two-photon polymerization (2PP) on silanized glass substrates using an UpNano NanoOne 1000 system (VAT mode) equipped with a 40\(\times\) and 1.4 NA oil-immersion objective. Prints were performed using the ``Standard Fine'' profile with a conservative slice mode, 0.525 \textmu{}m voxel height, 0.220 \textmu{}m voxel width, 0.250 \textmu{}m layer height, and a fine-line distance of 0.6 \textmu{}m at 10 mW laser power and 150 mm s\(^{-1}\) scan speed. The line direction was fixed at 0\(^\circ\), and a fine infill mode was used to generate freestanding grating lines within a 10000 \textmu{}m \(\times\) 5000 \textmu{}m \(\times\) 0.1 \textmu{}m print volume (70 \textmu{}m field of view). Post-print solvent development was carried out according to the manufacturer's material-handling protocol.

\hypertarget{single-surface-spin-coated-films}{%
\subsection{Single-surface spin-coated films}\label{single-surface-spin-coated-films}}

PDMS (Sylgard 184, Dow; 10:1 base: curing agent) was mixed, degassed, dispensed (1.0--1.2 mL) on Si--PR molds, spread by spin (500 rpm, 10--20 s), and cured (65 \(^\circ\)C, 6 h) in an ISO-7 cleanroom. Films (\(\sim\)500 \textmu{}m on 2-in molds) were peeled after cure. Curing at 65 \(^\circ\)C minimized PDMS shrinkage versus ambient cure.

\hypertarget{two-photon-fabricated-suction-cup-mold}{%
\subsection{Two-photon fabricated suction cup mold}\label{two-photon-fabricated-suction-cup-mold}}

Suction cup molds were fabricated by 2PP on 40 mm by 40 mm glass substrates using an UpNano NanoOne 1000 system (VAT mode) equipped with a 5x and 0.5 NA air immersion objective. The molds are fabricated as a base 26 000 \textmu{}m x 26 000 \textmu{}m x 100 \textmu{}m cube, followed by a 25 000 \textmu{}m x 25000 \textmu{}m x 400 \textmu{}m cube with 1000 \textmu{}m tall and 300 \textmu{}m thick walls going around the perimeter on top. ``Standard Coarse'' printing profile was used. Within these walls and on top of the surface of the 400 \textmu{}m cube, a 100 \textmu{}m thick cube was printed followed by a cylinder array of 100 \textmu{}m diameter and 100 \textmu{}m height with center-to-center spacing of 200 \textmu{}m in both X direction and Y direction, using the ``Standard Fine'' printing profile. Post-print solvent development was carried out according to the manufacturer's material-handling protocol.

\hypertarget{bulk-casting-of-dual-surface-films}{%
\subsection{Bulk casting of dual-surface films}\label{bulk-casting-of-dual-surface-films}}

The nanostructured silicon wafer was placed in the bottom of a flat-base glass dish. Degassed PDMS (Sylgard 184, Dow, 10:1 base:curing agent) was poured into the dish; an extra droplet was placed on the printed grating. The substrate carrying the 2PP processed structure was mounted on a fitting lid and was lowered to merge both volumes (Figure S1). Spacers were designed to ensure correct finished thickness. Casts were cured at room temperature (48 h) and at 65 \(^\circ\)C (6 h), observing no practical differences. After demolding films were cut to size. Bulk casting was not performed in the cleanroom. Molds were successfully reused to fabricate multiple devices using this method.

\hypertarget{nanocharacterization-sem}{%
\subsection{Nanocharacterization (SEM)}\label{nanocharacterization-sem}}

PDMS surface-relief meta-gratings were mounted on cross-section SEM stubs with copper tape, sputter-coated with 15 nm Pt/Pd (80/20) in a Cressington 208 HR B using planetary rotation with tilt and imaged on an FEI Apreo SEM with ETD detector in secondary-electron mode (Figure S2). Two-photon--fabricated gratings on glass were cleaned with 2-propanol, coated with 20 nm Au in a Luxor AU sputter coater, and imaged on a QUANTA 650 FEG SEM.

\hypertarget{optical-microscopy}{%
\subsection{Optical microscopy}\label{optical-microscopy}}

A Keyence VK-3000 3D laser scanning microscope was used to obtain differential interference contrast imaging and laser confocal surface scans. PDMS samples were cleaned with isopropyl alcohol before imaging. Geometry was imaged with 10x and 150x objectives. Samples were aligned using an in-mold registration mark.

\hypertarget{cie-color-mixing}{%
\subsection{CIE Color Mixing}\label{cie-color-mixing}}

Perceived colors of single and overlapping diffraction orders were computed in the CIE 1931 framework: for each wavelength at each diffracted order, we integrated the spectral power with the CIE 1931 color-matching functions to obtain XYZ, converted to chromaticity (x,y), and mapped to device values by a standard XYZ\(\rightarrow\)linear-sRGB transform followed by the canonical sRGB gamma correction; overlapping orders were combined additively using luminance weighting, with equal intensities assumed unless stated otherwise, so coincident orders (e.g., \(\lambda\) at m=1 and \(\lambda\)/2 at m=2) mix at the chromaticity midpoint. Simulated results and the computational process are shown in Figure S3.

\hypertarget{optical-power-measurement}{%
\subsection{Optical power measurement}\label{optical-power-measurement}}

Measuring the power of the diffracted orders from the grating is a way to characterize its optical response. Because the grating is intended to be measured in reflection, the power in each diffraction order is measured for the reflected orders as opposed to transmitted ones. Figure S4A shows the experimental setup. A 1 mW 532 nm laser was directed through a linear polarizer to ensure incident polarization perpendicular to the grating lines. The PDMS sample was mounted on a glass slide to maintain a flat surface. For each diffraction order, the reflected light was collected by a 40.0 mm focal length biconvex lens and focused on a Thorlabs S120VC Si photodiode power sensor. The polarizer was mounted 1 cm from the laser output and the sample at 28 cm from the polarizer. The biconvex lens and power sensor were moved manually in a radius of approximately 20 cm from the sample. Measurements of 1st, 2nd and 3rd orders were performed with the beam illuminating the sample at normal incidence. The 0th order and specular reflection from unpatterned PDMS were performed at an angle of 5 degrees. Measurements were made at three different spots per grating period and the average is taken. Results from the measurements are presented in Figure S4B--C. The power diffracted to each order is presented in terms of the relative efficiency \(\eta_{m}\  = \frac{P_{m}}{P_{0}}\ \) with \(P_{m}\) indicating the power in diffraction order m and \(P_{0}\) representing the power measured in specular reflection on unpatterned PDMS next to the gratings. The results verify the successful fabrication of PDMS gratings capable of diffracting light into multiple orders.

\hypertarget{phantom-validation}{%
\subsection{Phantom validation}\label{phantom-validation}}

To probe the operation and sensitivity of the resulting devices under controlled conditions we performed a benchtop phantom validation experiment (Figure S5). A fast-response piezoelectric pump (XP-S2-028, Lee Company) driven by a function generator imposes a periodic pressure waveform previously recorded from the radial artery. The wrist phantom consists of a soft tissue mimicking elastomer (Ecoflex 00-20, Smooth-On Corporation) cast around a 3D-printed radius and ulna (Figure S5E). This allows us to compress the radial artery against the underlying bone, simulating palpation and applanation tonometry. The OptoPatch prototype sits over the course of a radial artery analogue. Under a digital microscope, we extract pulsatile signals from the ``active'' coloration area under oblique illumination from an uncollimated broadband white light source. Tracking the ``active'' areas consistently results in high-quality signal extraction.

\hypertarget{in-vivo-recording}{%
\subsection{In vivo recording}\label{in-vivo-recording}}

13 participants were recruited from campus. All provided informed consent in advance. Continuous noninvasive blood pressure (cNIBP) was recorded using a Finapres type device (NIBP Nano INL382, ADInstruments), recorded at the intermediate phalanx of the middle finger on the same arm as the patch device was placed. Participants were asked to sit in a chair and place the instrumented arm resting on a table in front of them. Recording took place in two different rooms with different lighting configurations (Figure S12). Height correction was performed using HCU calibration following manufacturer instructions. Data was recorded and exported using LabChart 8 (ADInstruments). 60 FPS H.264-compressed video of the device was recorded via hand-held smartphone camera (Galaxy A55 5G, Samsung Electronics, n = 5; iPhone 14, Apple, n = 8) at distances between 15-30 cm. Video was obtained in each device's default ``Camera'' app, and exposure settings were left on ``Auto'' for each device. Simultaneous recordings were synchronized to cNIBP post-hoc using video metadata and sound markers, then aligned within the nearest cycle using cross-correlation. In a subset of participants, simultaneous recordings were also obtained using a piezoelectric fingertip pulse transducer (TN1012/ST, ADInstruments) and a consumer-grade green light reflectance fingertip PPG sensor selected for its continuous analog output capability. The PPG sensor was embedded in a 3D-printed plastic housing to shield it from ambient light.

\hypertarget{signal-processing}{%
\subsection{Signal processing}\label{signal-processing}}

Smartphone videos were digitally stabilized prior to processing in Matlab 2025a (The Mathworks). Regions of interest were assigned manually to cover approximately the same pixel range. For each frame, color information was averaged within the ROI separately for each channel. Resulting colorimetric time series were low-pass smoothed via convolution with the modified-sinc kernel of Schmid et al.~using kernel MS1, order = 4, window size = 10 samples at 60 Hz sample rate.\citep{ref43} Baseline wander was removed via subtraction of a low-pass filtered signal (order = 2, window size = 400 samples). The effective band-pass range was approximately 0.3-10.0 Hz.

\hypertarget{snr-calculation}{%
\subsection{SNR calculation}\label{snr-calculation}}

For the tensiometer data series, SNR was estimated by harmonic regression at the driven fundamental frequency. In 20 s sliding windows, sensor measurements were fit to a sinusoid at the drive frequency, allowing amplitude and phase to vary. The fitted fundamental component was taken as the coherent driven response, and the residual mean-square error was used as an estimate of non-coherent and non-fundamental variation. SNR was computed as the ratio of fitted fundamental power to residual power.

\hypertarget{coherence-and-correlation-measurements}{%
\subsection{Coherence and correlation measurements}\label{coherence-and-correlation-measurements}}

Signal coherence between channels was quantified using the magnitude-squared coherence (MSC), computed from Welch cross-/auto-spectral density estimates. Data were segmented into 1024-sample windows (FFT length = 1024) with 75\% overlap between adjacent segments, and spectra were averaged across segments to obtain a stable coherence estimate. Coherence was evaluated over the 0.5--8 Hz band. Beat-to-beat correlation was computed by segmenting the synchronized signals into individual beats using the cNIBP reference to determine fiducial points with a modified Pan-Tompkins algorithm. Individual beats were z-scaled and the Pearson correlation was calculated.

\hypertarget{statistical-considerations}{%
\subsection{Statistical considerations}\label{statistical-considerations}}

Multi-sensor comparison correlations were computed per-beat as Pearson's \emph{r} against the cNIBP reference. To conduct hypothesis testing on the correlation values, each \emph{r}-value was transformed using Fisher's z-transform, \(z = artanh(r)\). Per-participant mean \emph{z}-values were then computed for each modality and two-tailed paired Student \emph{t-}tests were run on the participant-level paired differences. Holm-Bonferroni correction was applied to correct for multiple comparisons. Effect size is reported as \(\Delta r\) after Fisher-\emph{z} averaging and back-transformation, \(\Delta r = \tanh\left( \bar{z}_{\mathrm{OptoPatch}} \right) - \tanh\left( \bar{z}_{\mathrm{Comparator}} \right).\) 95\% CI was estimated by paired participant bootstrap resampling with 10,000 resamples, preserving OptoPatch--comparator pairing. For each comparison, 8 participants were used, except for Opto-PPG which used 6 participants, excluding 2 PPG recordings due to technical error.

\hypertarget{generative-ai-usage}{%
\subsection{Generative AI usage}\label{generative-ai-usage}}

Generative AI tools (Codex (OpenAI), Claude Code (Anthropic)) were used to support code generation. All results were manually assessed and reviewed by the authors.

\FloatBarrier

\hypertarget{acknowledgements}{%
\section{Acknowledgements}\label{acknowledgements}}

We thank Dr.~Oddvar Uleberg for his input on this manuscript.

\hypertarget{funding}{%
\section{Funding}\label{funding}}

\begin{itemize}
\tightlist
\item
  Swiss National Science Foundation Project INC-META grant 10.004.197 (AX)
\item
  NTNU Rector AVIT Fund grant 949024104 (AX)
\item
  Research Council of Norway via support to the Norwegian Micro- and Nano-Fabrication Facility NorFab, grant 295864 (AX)
\item
  Central Norway Regional Health Authority Liaison Committee for Education, Research and Innovation grant 202230282 (NKS)
\end{itemize}

\hypertarget{author-contributions}{%
\section{Author Contributions}\label{author-contributions}}

\noindent\textbf{Conceptualization:} AX, TLS; \textbf{Methodology:} AX, TLS, VS, AGST, HV, JL; \textbf{Investigation:} TLS, VS, ETW, ESH, AGST, JL; \textbf{Fabrication:} VS, ESH, ETW, AGST; \textbf{Visualization:} TLS, AX, VS, AGST; \textbf{Resources:} AX, MS, NKS, HV; \textbf{Formal analysis:} TLS; \textbf{Data curation:} TLS, ETW; \textbf{Supervision:} AX, MS; \textbf{Writing---original draft:} TLS; \textbf{Writing---review \& editing:} TLS, AX, VS, ESH, ETW, JL, AGST, HV, MS, NKS.

\hypertarget{conflict-of-interest}{%
\section{Conflict of Interest}\label{conflict-of-interest}}

TLS, VS, MS and AX are named inventors on a patent filing relating to material in this manuscript. No other interests declared.

\hypertarget{data-availability-statement}{%
\section{Data Availability Statement}\label{data-availability-statement}}

The data that supports the findings of this study are available from the corresponding authors upon reasonable request.

\hypertarget{ethics-declaration}{%
\section{Ethics Declaration}\label{ethics-declaration}}

Human subject studies were conducted in accordance with the principles of the declaration of Helsinki. All participants provided informed consent in advance. Data collection was assessed by the Norwegian Agency for Shared Services in Education and Research, reference 275828.